\documentclass{aa}  

\usepackage{graphicx}
\usepackage{txfonts}
\usepackage{xcolor}
\usepackage[colorlinks, citecolor = blue]{hyperref}
\begin{document}

   \title{PRUSSIC II - ALMA imaging of dense-gas tracers in SDP.81}

   \subtitle{Evidence for low mechanical heating and a sub-solar metallicity in a z=3.04 dusty galaxy}

   \author{Matus Rybak\inst{1,2}
          \and J. van Marrewijk\inst{3,2}
          \and J. A. Hodge\inst{2}
          \and P. Andreani \inst{3}
          \and G. Calistro Rivera \inst{3}
          \and L. Graziani \inst{4,5}
          \and J. P. McKean \inst{6,7,8,9}
          \and S. Viti \inst{2}
          \and P. P. van der Werf\inst{2}
                }

   \institute{Faculty of Electrical Engineering, Mathematics and Computer Science, Delft University of Technology, Mekelweg 4, 2628 CD Delft, The Netherlands\\
         \email{m.rybak@tudelft.nl}
        \and Leiden Observatory, Leiden University, P.O. Box 9513, 2300 RA Leiden, The Netherlands 
        \and European Southern Observatory, Karl-Schwarzschild-Stra\ss e 2, 85748 Garching bei M\"unchen, Germany
        \and Dipartimento di Fisica, Universit\'a di Roma La Sapienza, Piazzale Aldo Moro 2, 00185 Roma, Italy
        \and INAF/Osservatorio Astronomico di Roma, Via di Frascati 33, 00040 Monte Porzio Catone, Italy
                \and ASTRON, Netherlands Institute for Radio Astronomy, Oude Hoogeveensedijk 4, NL-7991 PD Dwingeloo, the Netherlands
\and Kapteyn Astronomical Institute, University of Groningen, P.O. Box 800, 9700AV Groningen, the Netherlands
                \and South African Radio Astronomy Observatory (SARAO), P.O. Box 443, Krugersdorp 1740, South Africa
                \and Department of Physics, University of Pretoria, Lynnwood Road, Hatfield, Pretoria 0083, South Africa
             }

   \date{Received 2023 June 29; Revised 2023 August 4}

\abstract{
We present deep ALMA Band~3 observations of the HCN, HCO$^+$, and HNC (4--3) emission in SDP.81, a well-studied $z=3.042,$ strongly lensed galaxy. These lines trace the high-density gas, which remains almost entirely unexplored in $z\geq1$ galaxies. Additionally, these dense-gas tracers are potentially powerful diagnostics of the mechanical heating of the interstellar medium. While the HCN(4--3) and HNC(4--3) lines are not detected, the HCO$^+$(4--3) emission is clearly detected and resolved. This is the third detection of this line in a high-redshift star-forming galaxy. We find an unusually high HCO$^+$/HCN intensity ratio of $\geq2.2$. Based on the modelling of the photodissociation region, the most likely explanation for the elevated HCO$^+$/HCN ratio is that SDP.81 has low mechanical heating, making up less than 10\% of the total energy budget, along with a sub-solar metallicity of $Z\approx0.5$~$Z_\odot$. While such conditions might not be representative of the general population of high-redshift dusty galaxies, a lower-than-solar metallicity might significantly impact gas masses inferred from CO observations. In addition, we report the detection of CO(0--1) absorption from the foreground lensing galaxy and CO(1--0) emission from a massive companion to the lensing galaxy, approximately 50~kpc to the south-east. }

   \keywords{Galaxies: high-redshift -- Galaxies: individual: SDP.81 -- Galaxies: ISM -- Submillimeter: ISM
               }

   \maketitle
%

\section{Introduction}

One of the key questions in astrophysics is understanding the process of star formation, namely, exploring how cold, molecular gas is converted into newborn stars. Extensive surveys have revealed that the star-forming activity of the Universe peaks between redshifts $z=2-4$. Beyond $z\approx1$, cosmic star formation is dominated by dust-obscured, star-forming galaxies (DSFGs) with star formation rates (SFRs) of a few hundred to few thousand $M_\odot$/yr and far-infrared (FIR) luminosities of $\geq10^{12}$~$L_\odot$ \citep{Casey2014,Zavala2021}.

In this work, we  also consider how  the immense SFRs of DSFGs are linked to their molecular gas reservoirs. Current studies of cold gas in high-z DSFGs have focused mainly on the bright CO, [\ion{C}{ii}], and [\ion{C}{i}] emission lines (see reviews by \citealt{Carilli2013, Hodge2020}). However, these lines trace gas down to densities of $10^2$ to $10^3$~cm$^{-3}$, well below the typical densities of star-forming clouds. Crucially, they provide little to no insight into some of the key ingredients of star formation; namely, the high-density gas ($n\geq10^4$~cm$^{-3}$) and corresponding physical conditions. In particular, due to their intense SFRs, molecular clouds in DSFGs will be exposed to numerous shocks from supernovae, outflows from young stellar objects, and winds from massive stars injecting significant amounts of energy into the gas \citep{Loenen2008}. Indeed, mechanical heating can play a significant role in gas heating in nearby ultraluminous infrared galaxies (ULIRGs) as shown, for example, by \citet{Rosenberg2015}. 

Assessing the relative contribution of mechanical heating to the energy budget of DSFGs is challenging and limited to indirect probes. Recent studies of lensed DSFGs have found them to be overluminous in high-$J$ CO emission, which has been attributed to mechanical heating (\citealt{Riechers2021}, however, c.f. Butler et al., in prep.). Similarly, radiative transfer modelling of well-sampled CO ladders of \textit{Planck}-selected DSFGs has found significant non-thermal excitation \citep{Harrington2021}. However, using high-$J$ CO lines as mechanical heating tracers is fraught with significant uncertainties, as the high CO excitation can be caused by a number of diverse mechanisms, such as cosmic-ray or X-ray heating.

A more direct tracer of the dense gas and the associated physical conditions is high dipole moment molecules such as HCN, HCO$^+$, and HNC. Surveys of local galaxies have shown that the HCN(1--0) luminosity correlates linearly with the star-formation rate over eight orders of magnitude, from individual molecular clouds to entire galaxies (e.g. \citealt{Gao2004b, Wu2005, Bigiel2015, Jimenez2019}).
 Extending these $z\approx1$ studies to high redshift remains challenging, as the HCN emission can be more than 10$\times$ fainter than CO. Despite almost two decades of effort, only three $z\geq1$ DSFGs have been detected in HCN(1--0) (\citealt{Gao2007, Oteo2017, Rybak2022a}). Indeed, as  recently shown by \citet{Rybak2022a}, DSFGs might have low dense-gas fractions, making the HCN(1--0) emission even harder to detect.

An alternative to observing the ground-state transitions of HCN/HCO$^+$/HNC are the mid-$J$ transitions. The mid-$J$ lines are both intrinsically brighter and, at high redshift, conveniently fall into the easily accessible 3-mm atmospheric window \citep{Wagg2005}. The mid-$J$ lines have been proposed to be better tracers of dense gas than HCN(1--0) \citep{Krips2008, Viti2017}, which is often associated with densities much lower than its nominal critical density \citep{Kauffmann2017, Jones2023},
On the other hand, mid-$J$ transitions might be sensitive to, for example, mechanical heating or mid-infrared pumping of the vibrational modes \citep{Aalto2007,Kazandjian2012}. 

The sensitivity of the HCN/HCO$^+$/HNC lines to mechanical heating makes them potentially powerful tracers of turbulence in star-forming galaxies \citep{Loenen2008, Kazandjian2015}. For example, the HCN(4-3)/HCN(1-0) and HCN(4-3)/HCO$^+$ (4-3) ratios\footnote{Throughout this paper, the term `line ratios' refers to luminosity ratios of intensities in K km s$^{-1}$ pc$^2$ units, unless  otherwise specified.} increase by >1 dex between a relative mechanical heating of 0\% and 5\% \citep{Kazandjian2015}.

In local galaxies, the HCN/HCO$^+$/HNC (3--2) and (4--3) lines have been systematically targeted from the scales of entire galaxies and galactic centres \citep{Zhang2014, Israel2023} to kpc-scales \citep{Wilson2008, Tan2018, Jiang2020} and pc-scales (e.g. \citealt{Imanishi2018, Impellizzeri2019, Behrens2022}). However, at high redshift, the number of mid-$J$ detections remains very limited, with only a handful of sources detected in the (3--2), (4--3) or (5--4) emission lines \citep{Danielson2013, Oteo2017, Bethermin2018, Canameras2021} or in spectral stacking \citep{Spilker2014, Reuter2022, Hagimoto2023}.

To expand the number of high-z galaxies detected in dense-gas tracers, we launched \textsc{Prussic}\footnote{Prussic acid is another name for HCN, which was first isolated from the Prussian blue colour pigment.} -- a comprehensive census of dense-gas tracers in high-redshift star-forming galaxies. In the first \textsc{Prussic} paper, \citet{Rybak2022a} presented the Karl G. Jansky Very Large Array (JVLA) observations of the $J_\mathrm{upp}=1$ HCN, HCO$^+$, and HNC emission in six $z\sim3$ lensed DSFGs, finding low dense-gas fractions and elevated dense-gas star-forming efficiencies.

In this second paper of the \textsc{Prussic} series, we present deep, spatially resolved Atacama Large Millimeter / Sub-millimeter Array (ALMA) observations of the HCN/HCO$^+$/HNC(4--3) emission in SDP.81, a $z=3.042$ gravitationally lensed DSFG\footnote{We assume a flat $\Lambda$CDM cosmology, with $\Omega_m=0.315$ and $H_0=67.4$ km s$^{-1}$ Mpc$^{-1}$ \citep{Planck2016}.}. Using photodissociation region (PDR) modelling, we constrain the range of mechanical heating and metallicity in this galaxy. We also use the high fractional bandwidth of our observations to explore the environment of the foreground lensing galaxy.


\begin{table}
\caption{Global properties of SDP.81. Individual columns list the source position, source and lens redshift ($z_L$, $z_L$), sky-plane (observed) FIR and CO(1--0) luminosities, CO(1--0) FWHM \citep{Valtchanov2011}, and FIR continuum magnification \citep{Rybak2020b}. $L_\mathrm{FIR}$ is integrated over 8-1000~$\mu$m.}
    \centering
    \begin{tabular}{ll|c}
    \hline
    \multicolumn{3}{l}{Basic properties}\\
    \hline
     RA & [J2000] & 09:03:11.57  \\
     DEC & [J2000] & +00:39:06.5  \\
      $z_s$ & & 3.042\\
     $z_L$ & & 0.299 \\
     $L_\mathrm{FIR}^\mathrm{sky}$ & [$L_\odot$] & $(43\pm0.5)\times10^{12}$ \\
     $L{'}^\mathrm{sky}_\mathrm{CO(1-0)}$ & [K km s$^{-1}$ pc$^{2}$] & $(54\pm9)\times10^{10}$ \\
    FWHM$_\mathrm{CO(1-0)}$ &  [km s$^{-1}$] & 435$\pm$54 \\
    $\mu_\mathrm{160\mu m}$ & & 18.2$\pm$1.2\\
    
    \hline
        \multicolumn{3}{l}{ALMA Band 3 continuum}\\
        \hline

         $S^{\mathrm{sky}}_\mathrm{3\mathrm{mm}}$ & [mJy] & 0.510$\pm$0.031\\
       \hline
       
    \multicolumn{3}{l}{$J_\mathrm{upp}=1$ lines \citep{Rybak2022a}}\\
    \hline
     $L{'}^\mathrm{sky}_\mathrm{HCN(1-0)}$ & [K km s$^{-1}$ pc$^2$] & $\leq3.7\times10^{10}$ \\
    $L{'}^\mathrm{sky}_\mathrm{HCO^+(1-0)}$ & [K km s$^{-1}$ pc$^2$] & $\leq4.2\times10^{10}$ \\
    $L{'}^\mathrm{sky}_\mathrm{HNC+(1-0)}$ & [K km s$^{-1}$ pc$^2$] & $\leq2.4\times10^{10}$ \\
    
    \hline
    \multicolumn{3}{l}{$J_\mathrm{upp}=4$ lines (This work)}\\
    \hline

    $S\Delta v_\mathrm{HCN(4-3)}$ & [Jy km s$^{-1}$] & $\leq$0.302  \\
    $S\Delta v_\mathrm{HCO^+(4-3)}$ & [Jy km s$^{-1}$] & 0.357$\pm$0.056  \\
        $S\Delta v_\mathrm{HNC(4-3)}$ & [Jy km s$^{-1}$] & $\leq$0.207  \\
    \hline
          $L{'}^\mathrm{sky}_\mathrm{HCN(4-3)}$ & [K km s$^{-1}$ pc$^{2}$] & $\leq7.2\times10^9$ \\
    $L{'}^\mathrm{sky}_\mathrm{HCO^+(4-3)}$ & [K km s$^{-1}$ pc$^{2}$] & $(16.0\pm4.3)\times10^9$ \\
        $L{'}^\mathrm{sky}_\mathrm{HNC(4-3)}$ & [K km s$^{-1}$ pc$^{2}$] & $\leq7.1\times10^9$ \\

     \hline
    \end{tabular}
    \\
   \label{tab:sources}
\end{table}

\section{Observations and signal extraction}

\subsection{SDP.81:\ the prototypical lensed dusty galaxy}
SDP.81 is a $z=3.042$ DSFG, strongly gravitationally lensed by a foreground $z=0.299$ elliptical galaxy, with an inferred star-formation rate of $\approx$435~$M_\odot$ yr$^{-1}$ \citep{Rybak2020b}.
SDP.81 was initially identified in the H-ATLAS survey \citep{Negrello2010} and confirmed as being gravitationally lensed by \citep{Bussmann2013}. Detailed lens models have been derived from high-resolution ALMA sub-mm imaging \citep{Rybak2015a, Dye2015, Wong2015, Tamura2015, Hezaveh2016} and \textit{Hubble} Space Telescope (HST) near-infrared observations \citep{Dye2014}. SDP.81 was targeted by ALMA in Bands~4, 6, and 7 as part of the 2014 Long Baseline Campaign \citep{alma2015} and, more recently, in Bands~3 and 8 \citep{Rybak2020b}. These observations delivered high-resolution maps of the CO(5--4) and (8--7) \citep{Rybak2015b, Swinbank2015} and [\ion{C}{ii}] and CO(3--2) and (10--9) lines \citep{Rybak2020b}. Additional data include HST imaging \citep{Dye2014}, CO(1--0) observations \citep{Frayer2011, Valtchanov2011}, and \textit{Herschel} spectroscopy \citep{Valtchanov2011, Zhang2018}. 

High-resolution ALMA and HST imaging have revealed that SDP.81 consists of a compact, clumpy dusty disk $\approx$3~kpc in diameter \citep{Rybak2015a, Dye2015, Swinbank2015}, embedded in an extended CO and [\ion{C}{ii}] emission \citep{Swinbank2015, Rybak2015b, Rybak2020b}. The CO velocity fields show ordered rotation, but with significant perturbations, thus, SDP.81 has been classified as a post-coalescence merger \citep{Rybak2015b} and the CO-traced disk is Toomre-unstable \citep{Swinbank2015}. The merger scenario is further supported by the very extended [\ion{C}{ii}] \citep{Rybak2020b} and rest-frame UV emission, which form elongated, $\geq$15-kpc long tidal tails \citep{Dye2015, Hatsukade2015, Rybak2015b}. There is currently no indication of a buried AGN in SDP.81: the mid-infrared WISE and \textit{Spitzer} imaging \citet{Negrello2014} is consistent with pure star formation\footnote{While SDP.81 is detected in the XMM-Newton X-ray observations \citep{Ranalli2015}, this emission is likely associated with the bright AGN in the $z=0.299$ lensing galaxy.}. 

Most recently, \citet{Rybak2022a} targeted the HCN/HCO$^+$/HNC(1--0) emission in SDP.81 using deep JVLA imaging; none of the lines was detected. Table~\ref{tab:sources} summarises the main properties of SDP.81.

\subsection{ALMA Band~3 observations of SDP.81}

\begin{table*}
\caption{ALMA Band~3 observations of SDP.81: dates, time on-source, baseline ranges, precipitable water vapour (pwv), synthesised beam sizes (natural weighting), and sensitivity (measured over a 200~MHz bandwidth at the position of the HCN(4--3) line). \label{tab:alma} }
\begin{center}
 \begin{tabular}{@{}llccccc @{}}
 \hline \hline
ID & Dates observed & $t_\mathrm{on}$ & Baseline range & pwv & Beam FWHM & $\sigma_\mathrm{rms}^\mathrm{200\,MHz}$ \\
 & & [min] & [m] & [mm] & [arcsec] & [$\mu$Jy\,beam$^{-1}$ ] \\
 \hline
2016.1.00663.S  & 2017 September 9 & 46& 41 -- 7550 & 1.5 & 0.36$\times$0.24 & 126\\
2018.1.00747.S  & 2019 August 18, 27 & 129 & 28 -- 3640 & 0.8 -- 1.3 & 0.33$\times$0.29 & 35\\
 & 2021 July 24, 25; 2021 August 4  & 124 &  14 -- 3696 & 3.5 -- 3.9 & \multicolumn{2}{c}{Data unusable.}\\
2017.1.01694.S & 2018 April 25, 26 & 51 & 15 -- 500 & 5.5 -- 7.5 & 2.3$\times$2.1 & 51\\
\hline
 \hline
 \end{tabular}
\end{center}
\end{table*}

The ALMA Band~3 observations of SDP.81 consist of three different programmes, running from 2017 to 2021 and spanning multiple configurations. Table~\ref{tab:alma} lists individual observations. We now discuss the details of individual observing runs.

The project 2016.1.00663.S (Cycle~4, PI: Rybak) was carried out in a very extended configuration with baselines extending out to 7.5~km. The CO(3--2) line covered by this data was presented by \citet{Rybak2020b}. In addition to the CO(3--2) line, the observations covered the HCN(4-3) and HCO$^+$(4--3) emission, but not the HNC(4--3) line. 
The spectral setup consisted of two spectral windows (SPWs) centred at 85.73 and 87.68~GHz that were configured with a spectral resolution of 3.90625 and 7.81250~MHz, with a total width of 1.875 GHz each. The other two SPWs were centred at 97.72 and 99.72~GHz, with a resolution of 15.625~MHz and a total bandwidth of 2.0~GHz. 

The project 2018.1.00747.S (Cycle~6, PI: Rybak) was carried out in an extended configuration with baselines out to 3.6~km. The observations were taken in two batches, one in the summer of 2019 and another in the summer of 2021. The spectral setup was the same as above. 

Unfortunately, the 2021 data were taken in bad weather conditions with copious wet clouds. The data quality was marginal; the phase calibration could not be performed successfully, and the 'check' sources were not detected. Consequently, we have excluded the 2021 observations from our analysis.

Observations for project 2017.1.01694 (PI: Oteo) were taken in a compact configuration, with a maximum baseline length of 500~m. The spectral setup differed from the previous two programmes\textbf{: namely, it used} two SPWs centred at 87.988 GHz and 89.688~GHz, with a resolution of 15.625~MHz. This is the only setup that covers the HNC(4--3).

The data were reduced using the standard ALMA pipeline and \textsc{Casa} versions 4.7 and 5.4 \citep{McMullin2007}. After concatenating all the data, we re-calculate the noise on individual visibilities using \textsc{casa}'s \texttt{statwt} task; this ensures that the noise is estimated consistently for all the scheduling blocks. For the frequency range covering the HCN(4--3) and HCO$^+$(4--3) lines, the resulting dataset totals 226~min on-source and provides sensitivity to spatial scales between 0.29 and 46~arcsec at 88.7~GHz. The HNC(4--3) line is only covered by the compact-array observations with 51~min on-source; the array configuration provides sensitivity to spatial scales between 2.1 and 46~arcsec at 88.7~GHz. The primary beam FWHM was 65~arcsec.

\subsection{Imaging}

\begin{figure}[h]
\centering
    \includegraphics[height=3.3cm]{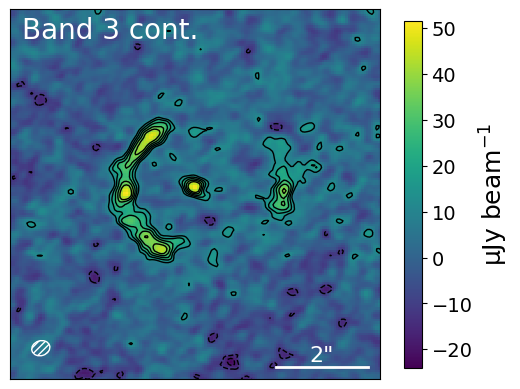}
    \includegraphics[height=3.3cm]{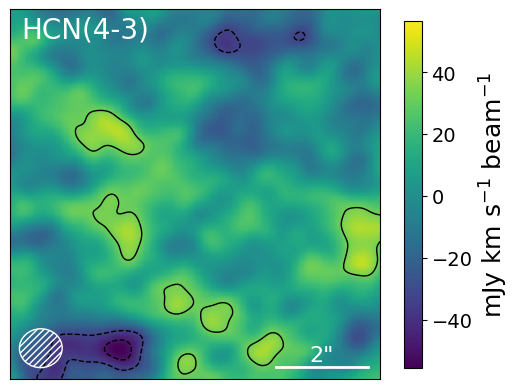}\\
    \includegraphics[height=3.3cm]{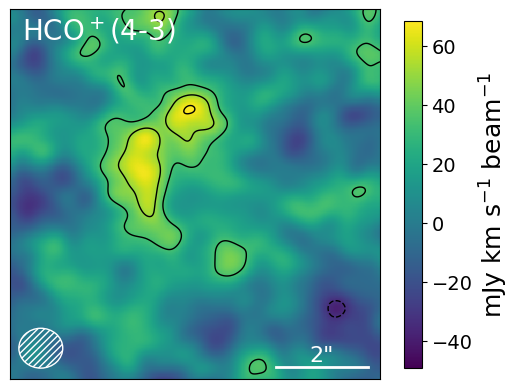}
    \includegraphics[height=3.3cm]{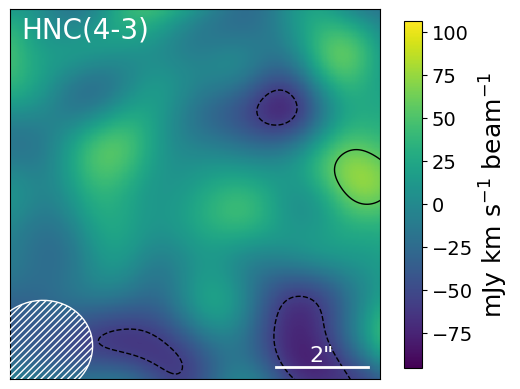}\\

    \caption{ALMA narrow-band images of SDP.81. \textit{Upper left}: Rest-frame 847-$\mu$m continuum at native resolution; the main Einstein arc and the counter-image are clearly visible. The point-source emission in the centre of the image is from the previously identified AGN in the lensing galaxy at $z=0.299$. \textit{Upper right and lower panels}: Narrow-band images at the systemic frequencies of the HCN(4--3), HCO$^+$(4--3), and HNC(4--3) lines. The images are collapsed over 430 km~s$^{-1}$ bandwidth. Contours start at $\pm2\sigma$, with a 1$\sigma$ increment. The significantly larger beam in the HNC image is caused by the lack of long-baseline observations at this frequency. The HCN(4--3) and HNC(4--3) lines are not detected, but HCO$^+$(4--3) is clearly detected and resolved.}
    \label{fig:alma_hcnhcohnc}
\end{figure}

To image the Band~3 continuum and the three emission lines, we used \textsc{Casa}'s \texttt{tclean} task. We first imaged the continuum using the line-free channels (86.0-88.5~GHz and 90.0-101.0~GHz) and a manually drawn mask, cleaning down to 1.5$\sigma$ (Fig.~\ref{fig:alma_hcnhcohnc}). We set the \texttt{tclean}'s parameter \texttt{fastnoise=False} to properly re-calculate the noise per baseline and channel.

\begin{figure}
\centering
    \includegraphics[width=0.49\textwidth]{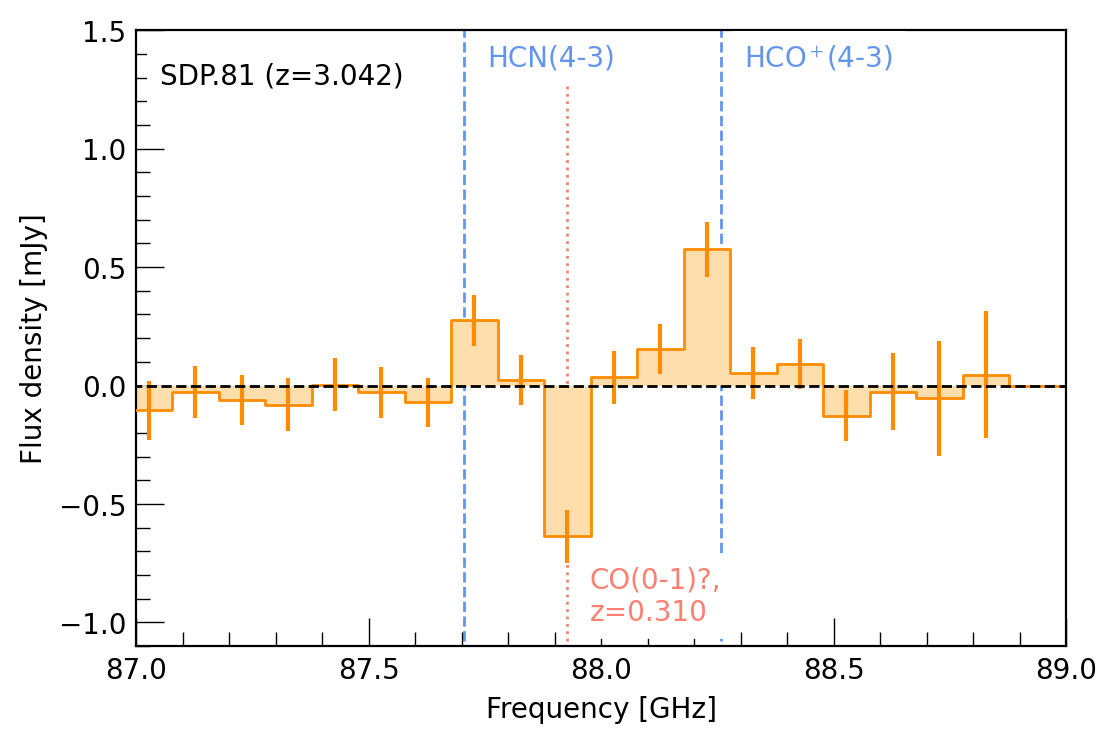}    
    \caption{ALMA Band~3 spectrum of SDP.81, extracted from the main Einstein arc. We derive the spectrum from dirty-image cubes with a 1-arcsec taper and a spectral resolution of 100~MHz. There is a clear positive excess at the position of the HCO$^+$(4--3) line, as well as potential foreground CO(0--1) absorption (see Appendix A.1). Note: the line fluxes in Table~\ref{tab:sources} are extracted from the narrow-band images (Fig.~\ref{fig:alma_hcnhcohnc}).}
    \label{fig:alma_spectrum}
\end{figure}

To image the dense-gas tracers, we subtracted the continuum emission using the \texttt{uvcontsub} task, fitting a constant flux to the line-free part of the spectrum. Due to the faintness of the lines, we produced dirty images only (i.e. without any deconvolution). For the HCN(4--3) and HCO$^+$(4--3) lines, the combination of discrepant array configurations produces a dirty beam with very strong sidelobes; we mitigate this effect by using different $uv$-plane tapers with a final beam FWHM of $\sim$0.85''$\times$0.94''. No taper was applied to the HNC(4--3) data as it was taken with a compact configuration. To examine the data in the spectral dimension, we created dirty-image cube of the entire dataset. Figure~\ref{fig:alma_spectrum} shows the resulting spectrum extracted from the main lensing arc.

Finally, we made `narrow-band' (moment-0) images centred on the expected line frequency, choosing a bandwidth of 430 km s$^{-1}$, corresponding to the CO(3--2) FWHM. 
At $z=3.042$, the rest-frame (observed-frame) frequencies of the lines are: HCN(4--3): 354.506~GHz (87.706~GHz); HCO$^+$(4--3): 356.734~GHz (88.257~GHz); and HNC(4--3): 362.630~GHz (89.716~GHz).  These are presented in Fig.~\ref{fig:alma_hcnhcohnc}. The resulting $\sigma_\mathrm{rms}$ levels are: HCN(4--3): 39 $\mu$Jy/beam, HCO$^+$(4--3): 38~$\mu$Jy/beam, HNC(4--3): 66~$\mu$Jy/beam. Finally, in addition to the line and continuum emission from the background DSFG, we also report the detection of a foreground CO(0--1) absorption and of a gas-rich companion to the lensing galaxy - see Appendix~\ref{app_A} for more details.

\section{Results}
\label{sec:results}

\subsection{HCN, HCO$^+$, and HNC emission}

As shown in Fig.~\ref{fig:alma_hcnhcohnc}, the HCN(4--3) and HNC(4--3) lines are undetected in the dirty-image maps. On the contrary, the HCO$^+$(4--3) line is clearly detected and resolved: the peak signal-to-noise ratio (S/N) is 4.3$\sigma$ and we see $\geq3\sigma$ excess emission over 4--5 beams.  The HCO$^+$(4--3) emission is located at the centre and north of the main Einstein arc; the southern part of the arc and the counter-image (to the West) are not detected. This is only the third detection of the HCO$^+$(4--3) in a high-redshift DSFG, after ACT J2029+0120 ($z=2.64$, \citealt{Roberts2017}) and G244.8+54.9 ($z=3.01$, \citealt{Canameras2021}); additionally, HCO$^+$(4--3) was detected in the Cloverleaf quasar ($z=2.56$, \citealt{Riechers2011b}). Compared to the dust continuum, the HCO$^+$(4--3) emission seems to be concentrated to the north; however, given the low S/N of our data, the HCO$^+$(4--3) morphology can not be reliably assessed. The low S/N also precludes a reliable reconstruction of the HCO$^+$(4--3) emission in the source plane.

We extracted the HCO$^+$(4--3) line flux and upper limits for HCN(4--3) and HNC(4--3) from a manually drawn aperture over the main lens arc (note:\ the contribution from the counterimage is negligible). The aperture area corresponds to $\approx$6 beams for HCN and HCO$^+$, and 1.5 beams for HNC. For HCN and HNC fluxes, we adopt the corresponding $3\sigma$ upper limits. To calculate the line luminosities, we multiplied the measured fluxes by a factor of 1.31 (a ratio between the total line flux and the flux contained within the line FWHM). Table~\ref{tab:sources} lists the derived sky-plane luminosities.

\begin{figure}[h]
\centering
     \includegraphics[height=6cm]{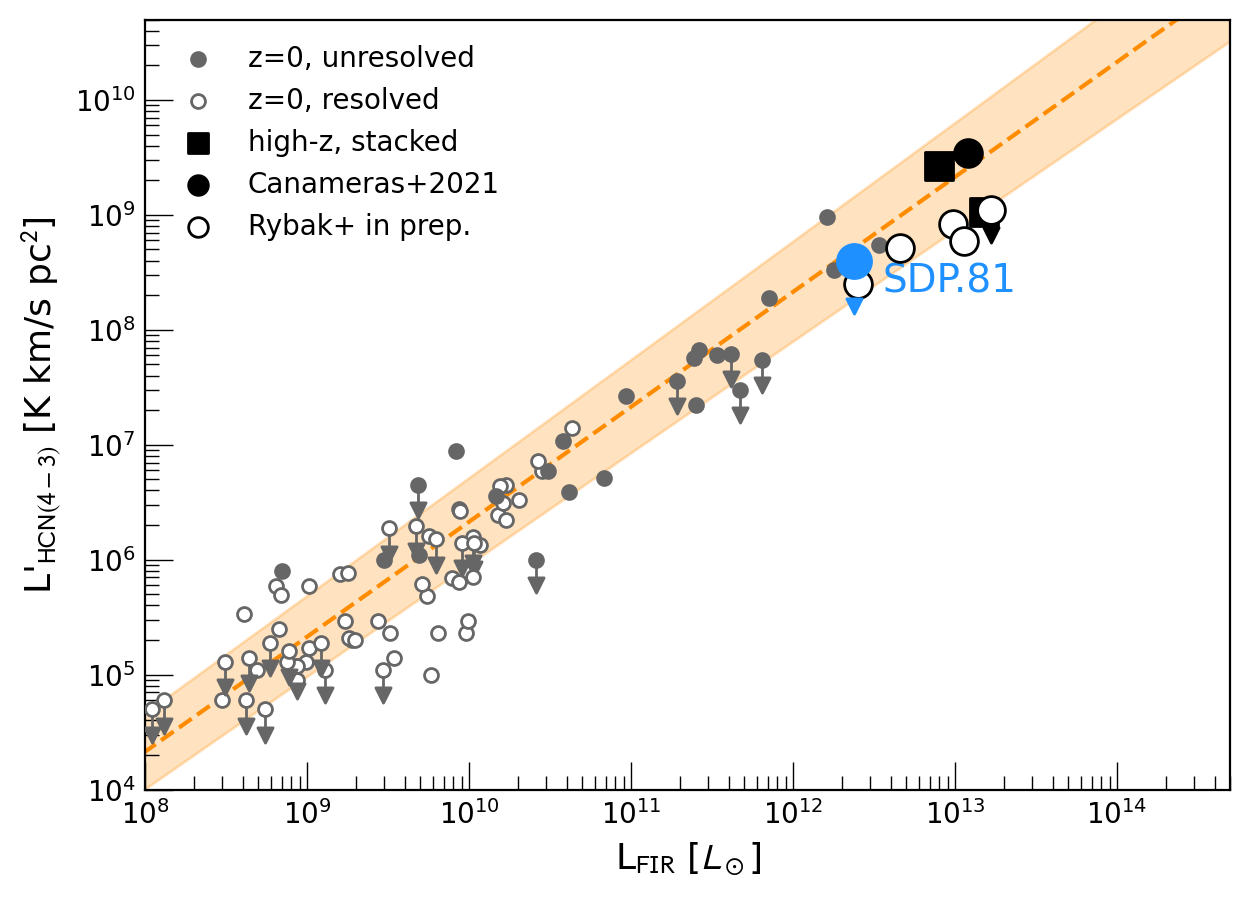}
    \includegraphics[height=6cm]{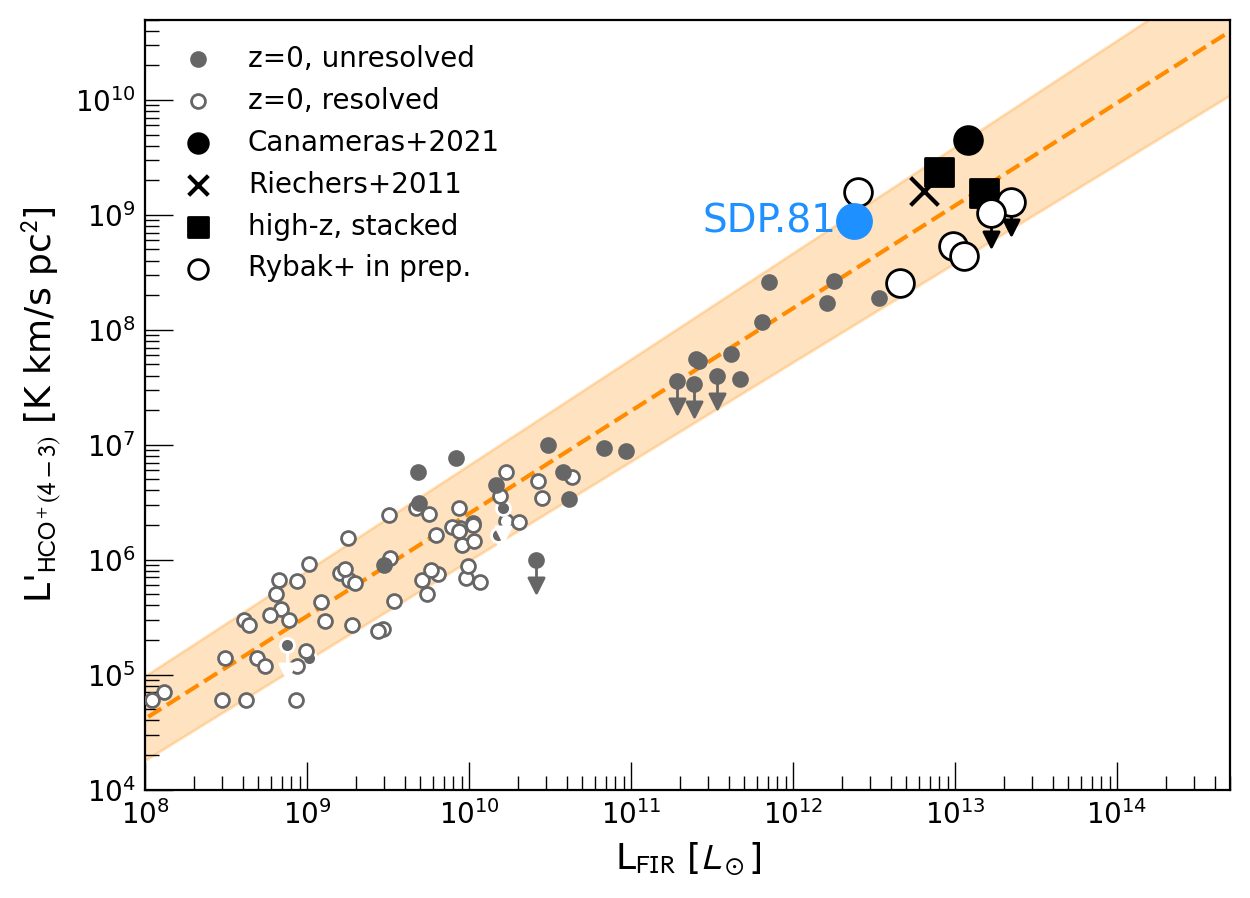}
    
    \caption{Compilation of HCN(4--3) vs far-IR luminosity (upper) and HCO$^+$(4--3) vs far-IR luminosity (lower) observations. Individual datapoints show $z=0$ galaxy-averaged \citep[grey filled points]{Greve2009, Papadopoulos2014, Zhang2014} and resolved observations \citep[grey open points]{Tan2018}, and high-$z$ detections in individual galaxies \citep[black, solid]{Riechers2011, Canameras2021} and Rybak et al., in prep. (black, open points) and spectral stacks \citep{Spilker2014, Hagimoto2023}. Where appropriate, luminosities are corrected for the lensing magnification. Dashed lines indicate the empirical trends from \citet{Zhang2014}. The error bars for the HCO$^+$(4--3) flux in SDP.81 are smaller than the data point (S/N$\approx$4). SDP.81 is $\approx$1$\sigma$ above the mean \citet{Zhang2014} HCO$^+$(4--3)-FIR trend.}
    \label{fig:corr_fir_hcn_hco}
\end{figure}

Figure~\ref{fig:corr_fir_hcn_hco} puts our HCN(4--3) and HCO$^+$(4--3) measurements in the context of $z=0$ studies \citep{Zhang2014, Tan2018} and the predicted $L_\mathrm{FIR}-L'$ relations from \citet{Zhang2014}\footnote{Note: the correct \citet{Zhang2014} $L_\mathrm{FIR}$ - $L'_\mathrm{HCO^+}$ correlation should be $\log L_\mathrm{FIR} = (1.12\pm0.05) \times \log L'_\mathrm{HCO^+(4-3)} + 2.83\pm0.34$ (Z.\,Zhang, priv.\,comm.). }. The HCO$^+$(4--3) is $\approx$1$\sigma$ above the local trend, whereas the HCN(4--3) upper limit falls close to the \citet{Zhang2014} trend. We do not show a similar plot for the HNC(4--3) line due to the limited number of detections at $z\sim0$.

\begin{figure}
\centering
    \includegraphics[width=0.45\textwidth]{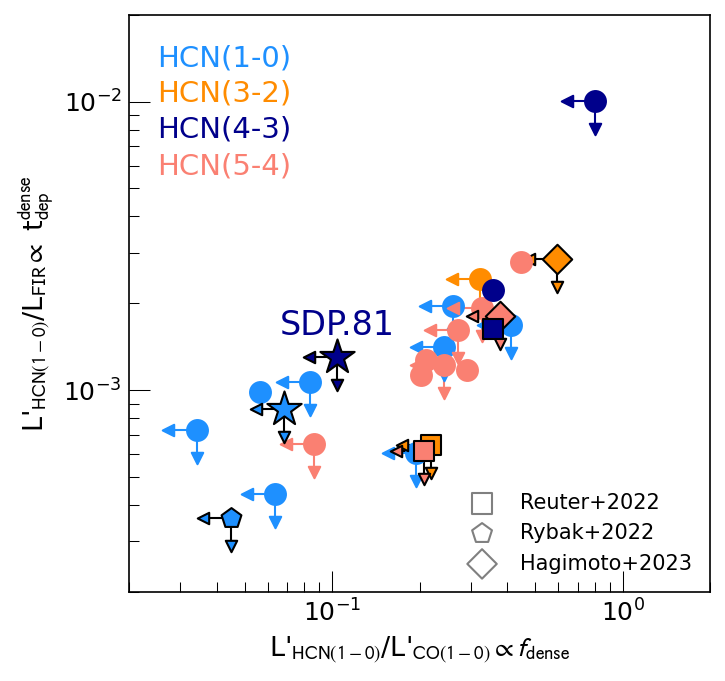}    
    \caption{Comparison of $L'_\mathrm{HCN}$/$L'_\mathrm{CO(1-0)}$ and $L'_\mathrm{HCN}$/$L_\mathrm{FIR}$ ratios for high-redshift galaxies, with SDP.81 datapoints denoted by stars. We include measurements for individual galaxies \citep[bullet points]{Danielson2013, Oteo2017, Bethermin2018,Canameras2021,Rybak2022a} and stacks \citep{Reuter2022,Rybak2022a,Hagimoto2023}. We convert the $J_\mathrm{upp}\geq2$ luminosities to HCN(1--0) following the $z\sim0$ HCN ladder from \citet{Israel2023}. The upper limits on HCN(4--3) emission in SDP.81 are consistent with recent works indicating low HCN/CO and HCN/FIR ratios in DSFGs.}
    \label{fig:sfe_fdense}
\end{figure}

The HCN/CO and HCN/FIR luminosity ratios can be used as rough proxies for the dense-gas fraction and dense-gas star-forming efficiency (e.g. \citealt{Gao2004b,Usero2015,Jimenez2019}, but c.f. \citealt{Israel2023}). Figure~\ref{fig:sfe_fdense} shows our HCN(4--3) measurements in SDP.81 in the context of other high-redshift observations of $J_\mathrm{upp}$=1, 3, 4, and 5 lines. For a better comparison, we converted the $J_\mathrm{upp}\geq2$ luminosities using the HCN ladder\footnote{Specifically, we consider the nine galaxies from \citet{Israel2023} which are detected in all HCN $J_\mathrm{upp}=1-5$ lines.} from the compilation of \citet{Israel2023}; the corresponding mean line ratios are$L'_\mathrm{HCN(3-2)}/L'_\mathrm{HCN(1-0)}=0.322$, $L'_\mathrm{HCN(4-3)}/L'_\mathrm{HCN(1-0)}=0.128$. The HCN(5--4) line is not included in the \citet{Israel2023} data; following \citet{Bethermin2018}, we use radiative transfer predictions for the Milky Way's central molecular zone from \citet{Mills2013} to set $L'_\mathrm{HCN(5-4)}/L'_\mathrm{HCN(1-0)}=0.094$.

We can put rough constraints on the HCO$^+$ excitation by combining our HCO$^+$(4--3) detection with the HCO$^+$(1--0) non-detection from \citet[see values in Table~\ref{tab:sources}]{Rybak2022a}. The $L'_\mathrm{HCO^+(4-3)}/L'_\mathrm{HCO^+(1-0)}$ ratio is $\geq$0.38, that is, consistent with a sub-thermal excitation. This lower limit on the $L'_\mathrm{HCO^+(4-3)}/L'_\mathrm{HCO^+(1-0)}$ ratio is SDP.81 is comparable to the mean HCO$^+$(4--3)/(1--0) ratio from the \citet{Israel2023} compilation of nearby galaxies (0.31$\pm$0.15).

\begin{figure}
\centering
     \includegraphics[width=0.45\textwidth]{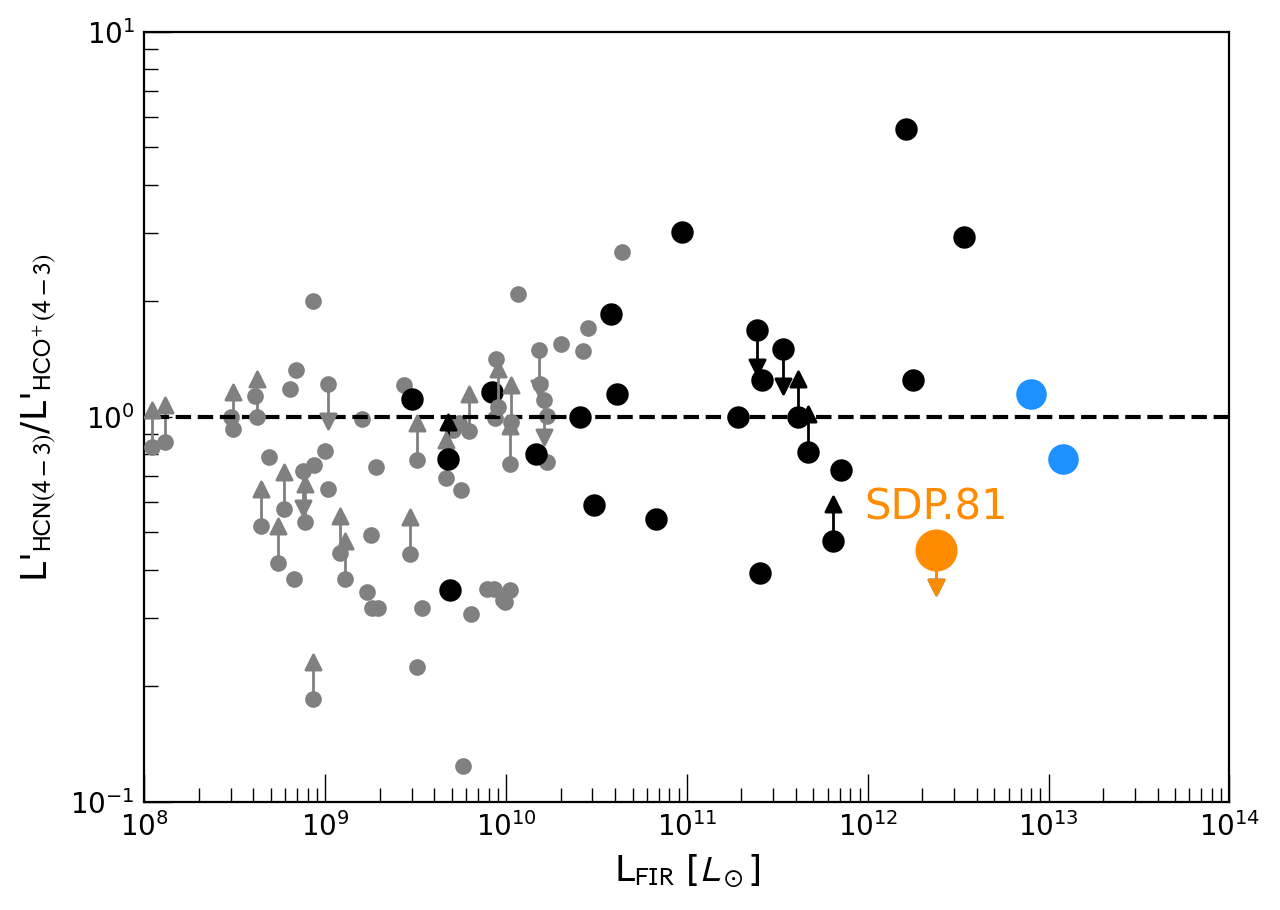}

    \caption{Observed HCN(4--3)/HCO$^+$(4--3) ratios as a function of far-IR luminosity. In grey: resolved measurements in $z\sim0$ galaxies from \citet{Tan2018}; black: galaxy-averaged $z\sim0$ measurements from \citet{Zhang2014}; blue: high-z galaxies. The HCN/HCO$^+$ ratio in SDP.81 is lower than in most extragalactic sources (for both low- and high-redshift ones).}
    \label{fig:fir_hcn_hco}
\end{figure}

As shown in Fig.~\ref{fig:fir_hcn_hco}, the slightly overluminous HCO$^+$(4--3) line and the upper limit on HCN(4--3) combine to a HCO$^+$(4--3)/HCN(4--3) ratio of $\geq$2.2. This value is significantly higher than galaxy-averaged values for nearby galaxies from \citet{Zhang2014}, although some individual regions in nearby galaxies show HCO$^+$/HCN(4--3) ratios up to $\sim$10 \citep{Tan2018, Galametz2020} on $\leq$10-pc scales. Looking at the ground-state transitions, HCO$^+$/HCN(1--0) ratios in nearby galaxies range between 0.4 and 2.5 \citep{Gracia2006,Garcia2012,Privon2015}. The HCO$^+$/HCN ratio tends to be slightly higher in purely star-forming galaxies compared to the AGN hosts \citep{Privon2015}, although the exact link with the AGN activity remains unclear \citep{Privon2020}.

While different spatial distributions of the tracers might result in differential magnification, we consider this effect to be limited: HCO$^+$ and HCN are expected to be co-spatial on kpc-scales and previous modelling of dust continuum and CO and [\ion{C}{ii}] emission in SDP.81 found magnification factors varying by $\leq$15\% \citep{Rybak2015a, Rybak2015b, Swinbank2015, Rybak2020b}, which are too small to explain the observed high HCO$^+$/HCN ratio.

\subsection{Photon-dissociation region modelling}

To investigate which physical conditions cause the high HCO$^+$/HCN(4--3) ratio, we used photon-dissociation region (PDR) modelling. Specifically, we explored the impact of varying mechanical heating and gas-phase metallicity
\footnote{For the solar metallicity, we assume the \citet{Asplund2009} value of $12+\log(\mathrm{O/H})=8.69$.}.

We adopted the PDR models of \citet{Kazandjian2012, Kazandjian2015} which are an extension of the \textsc{Leiden PDR-XDR} models of \citet{Meijerink2005}. These PDR models assume a semi-infinite parallel-plane slab morphology illuminated by external UV radiation, alongside X-ray and cosmic ray (CR) contributions. The models span a wide range of metallicity (0.1-2.0~$Z_\odot$) and mechanical heating\footnote{We also explore the option of increasing the CR ionisation rate above the default value; models with extra CR contribution are ruled out as they overpredict the HCO$^+$(4--3)/HCN(1--0) ratio.}. The latter is parametrised by a factor $\alpha$, the ratio of the mechanical and photoelectric heating at the surface of the cloud. Therefore, $\alpha=0$ corresponds to no mechanical heating, while $\alpha=1$ implies that mechanical heating is equal to photoelectric heating. The incident radiation field spectral energy distribution and the corresponding mechanical heating are derived assuming a Salpeter stellar initial mass function. The mechanical heating is assumed to be due to supernovae shock dissipation only and is implemented following the \citet{Loenen2008} prescription, neglecting the contributions from young stellar objects (which are very short-lived) and stellar winds (which contribute $\leq$6\% of the total mechanical heating, \citealt{Kazandjian2012}); the energy is injected uniformly throughout the cloud volume. Given the lack of evidence for an AGN in SDP.81, we assume all radiation is due to star formation.

\begin{figure*}[t]
\centering

     \includegraphics[height=6cm]{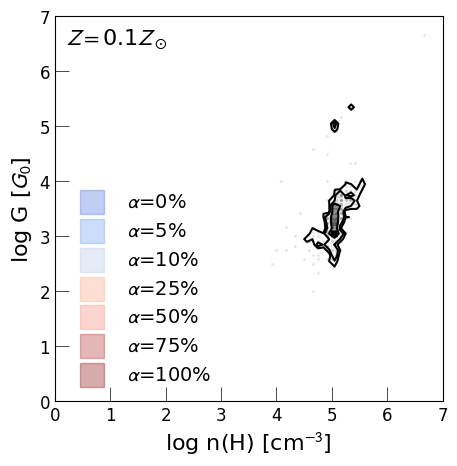}
     \includegraphics[height=6cm]{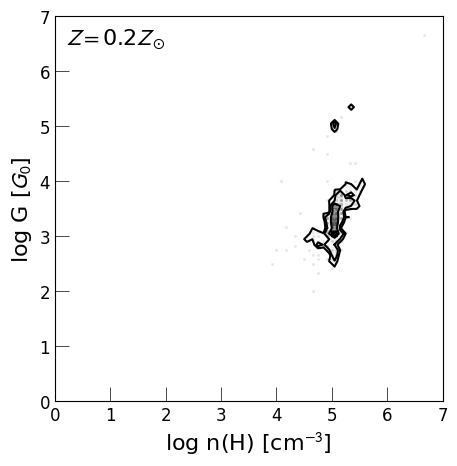}
     \includegraphics[height=6cm]{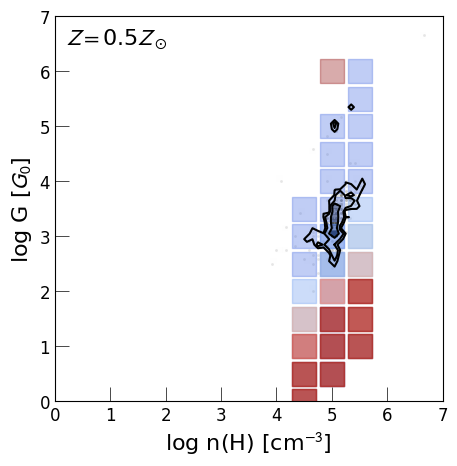}\\
     \includegraphics[height=6cm]{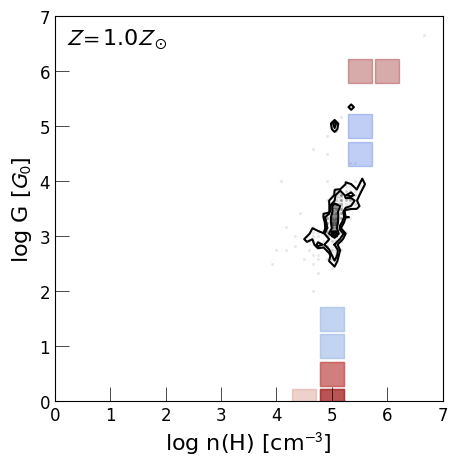}
     \includegraphics[height=6cm]{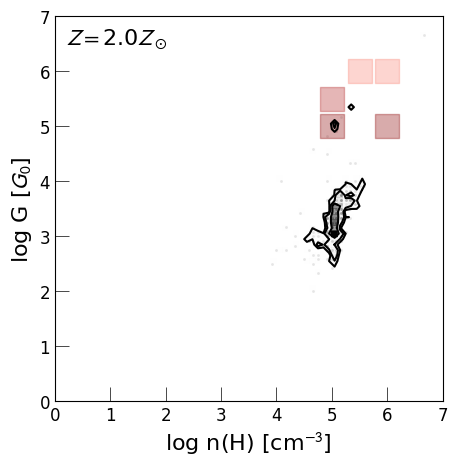}
     \includegraphics[height=6cm]{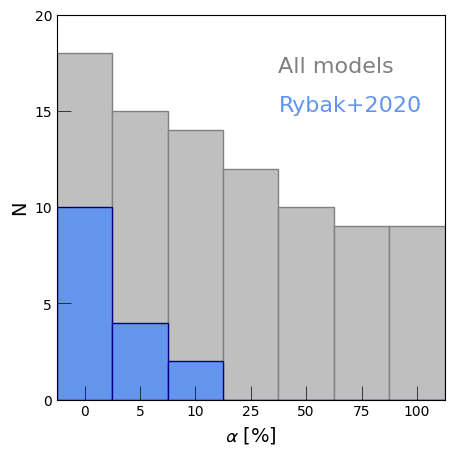}\\
    \caption{Far-UV radiation ($G$) and gas density ($n$) from PDR models, with constraints from the HCN, HCO$^+$, HNC, and CO(5--4) observations. Individual panels show models for $Z$=0.1, 0.2, 0.5, 1.0, 2.0~$Z_\odot$. Different colours denote the different levels of mechanical heating contribution $\alpha$ (0\% to 100\%); the coloured squares denote models consistent with the observed line ratios within 50\%; grey shaded contours show the $G$ and $n$ inferred from high-resolution imaging of SDP.81 by \citet{Rybak2020b}. Only the $Z$=0.5~$Z_\odot$, $\alpha$=0, 10\% models are consistent with the data. The most direct interpretation is that SDP.81 has sub-solar metallicity and only limited mechanical heating. \textit{Lower right:} Histogram of mechanical heating factor $\alpha$ for the $Z=0.5$~$Z_\odot$ model. All PDR models consistent with the line data are shown in grey; the models consistent with $G$ and $n$ inferred from \citet{Rybak2020b} are highlighted in blue.}
    \label{fig:pdrs}
\end{figure*}

For the comparison with models, we used the following dense-gas line ratios: $L'_\mathrm{HCO^+(4-3)}$/$L'_\mathrm{HCN(4-3)}\geq2.2$, $L'_\mathrm{HCO^+(4-3)}$/$L'_\mathrm{HNC(4-3)}\geq2.3$, and $L'_\mathrm{HCO^+(4-3)}$/$L'_\mathrm{HCN(1-0)}\geq3.8$ (see Table~\ref{tab:sources}). For further constraints, we also included  the CO(5--4) line, which is shown to closely trace the dust continuum emission \citep{Rybak2015b, Rybak2020b}. We did not include the CO(8--7) and (10--9) lines, which are concentrated in the northern part of the sources, and the CO(3--2), whose S/N value is too low to be useful for PDR modelling. We conservatively assume 50\% errors on the line ratios to account for the flux calibration uncertainties and potential spatial offsets between different tracers.

Figure~\ref{fig:pdrs} shows the regions in the $G$-$n$ space that are consistent with these constraints. The models with $Z=0.1$~$Z_\odot$ and $2.0$~$Z_\odot$ are effectively excluded by the observations. Models with $Z=1$~$Z_\odot$ are consistent with the data for $\alpha=$5 -- 10\%, but imply unrealistically low gas densities ($\leq10^2$~cm$^{-3}$) that are ruled out by the existing mid- and high-$J$ CO detections.

However, lowering the metallicity to $Z=0.5$~$Z_\odot$ allows the data to be reproduced over a wide range of $G$, $n$, and $\alpha$. Lowering the metallicity even further to $Z=0.2$~$Z_\odot$ and 0.1~$Z_\odot$ does not yield any physical solutions. In the lower right panel of Fig.~\ref{fig:pdrs}, we show the histogram of all $\alpha$ for all $Z=0.5$~$Z_\odot$ models consistent with the observed dense-gas tracers line ratios (grey). If we restrict the parameter space to the $G$, $n$ range from the 200-pc resolution PDR modelling from \citet{Rybak2020b}, only models with low mechanical heating ($\alpha$=0--10\%) are left. When we repeated our analysis including the CO(8--7) line (which is concentrated in the northern part of the source), the only feasible models are $Z=0.5$~$Z_\odot$ and $\alpha=0-5$\%. We therefore conclude that the most direct explanation of the high HCO$^+$/HCN ratio in SDP.81 is a combination of a lower-than-solar metallicity and low (or outright negligible) mechanical heating.

\section{Discussion}

\subsection{Mechanical heating in high-z dusty star-forming galaxies}

Several recent studies have argued for a significant mechanical heating in DSFGs due to large cloud-scale turbulence. In particular, \citet{Riechers2021} found that high-z DSFGs are overluminous in CO(9--8) emission compared to $z\sim0$ empirical trends and argued that this excess luminosity is due to mechanical heating. \citet{Harrington2021} claimed the presence of significant mechanical heating in lensed \textit{Planck}-selected, starburst DSFGs, on the basis of radiative transfer modelling of CO ladders. However, mechanical heating is not the only mechanism that can produce highly excited CO ladders and other large-sample studies have not found DSFGs to be overluminous in CO(9--8) (K. Butler, priv. comm.).

Our PDR modelling shows that the HCO$^+$/HCN/HNC emission in SDP.81 is consistent with little to no mechanical heating. A similar conclusion for this source was reached by \citet{Rybak2020b} based on the analysis of dust, [\ion{C}{ii}], and CO (3--2) to (10--9) lines. We note that unlike the \citet{Riechers2021} sample, SDP.81 is not overluminous in the high-J CO emission and its $L_\mathrm{FIR}-L'_\mathrm{CO(10-9)}$ ratio is consistent with local star-forming galaxies (\citealt{Liu2015}, see also \citealt{Kamenetzky2015}). Also, compared to the \citet{Harrington2021} sample, which is composed of the most extreme DSFGS with SFRs of few 1000~$M_\odot$/yr, SPD.81 has a SFR of 400~$M_\odot$/yr, and might be  inherently less turbulent as a result. Nevertheless, radiative transfer analysis indicates that the HCN and HCO$^+$ ladders are much more sensitive to mechanical heating than the CO emission \citep{Kazandjian2015}. Future studies of the HCN, HCO$^+$, and HNC ladders in large samples of high-redshift galaxies will be necessary to properly assess the role of mechanical heating in their ISM thermodynamics and how it varies with galaxy properties.

\subsection{Metallicity estimates for high-z dusty galaxies}

Gas-phase metallicity measurements in DSFGs are very sparse, as the widely-used rest-frame optical and near-infrared indicators suffer from extreme dust obscuration \citep{Maiolino2019}. Consequently, a different set of tracers (unaffected by the dust) is necessary to measure the metallicity of obscured, sub-mm bright galaxies.

One option is to use emission lines in the rest-frame far-infrared spectrum. The potential of far-infrared lines as a metallicity tracer was first studied by \citet{Nagao2011}, who used radiative transfer models of \ion{H}{ii} regions to predict the dependence on different fine-structure lines on the gas-phase metallicity. \citet{Nagao2011} found that the combination of the [\ion{O}{iii}] 52- and 88-$\mu$m lines and the [\ion{N}{iii}] 57-$\mu$m line provides good metallicity diagnostics. Similar conclusions were reached by \citet{Pereira2017}, who also used the \textit{Herschel} observations of O and N fine-structure lines to measure metallicity in a sample of highly obscured $z\sim0$ ULIRGs, obtaining values between $Z=0.7-1.5$~$Z_\odot$.

At high redshift, \citet{Wardlow2017} used \textit{Herschel} observations of the [\ion{O}{iii}] 52-$\mu$m and [\ion{N}{iii}] 57-$\mu$m lines to infer a metallicity of $\geq$1~$Z_\odot$ for a stacked spectrum of 13 lensed DSFGs\footnote{The only source in the \citet{Wardlow2017} sample that is directly detected in the [\ion{O}{iii}] 52-$\mu$m emission -- NGP NA.144 -- provides only an upper limit of a few $Z_\odot$ and is consistent with a sub-solar metallicity.}. Alternatively, \citet{Rigopoulou2018} used the [\ion{O}{iii}] 88-$\mu$m / [\ion{N}{ii}] 122-$\mu$m line ratios in combination with far-IR flux ratios to estimate the gas-phase metallicity in three $z=2-3$ galaxies. These latter authors inferred metallicity ranges of $Z=0.6-1.0$~$Z_\odot$ for HLSW01 ($z=2.96$) and $Z=0.7-1.1\,Z_\odot$ for J02399 ($z=2.80$). Their third source is SDP.81, for which they obtained only a weak upper limit $Z\leq2$~$Z_\odot$, consistent with our value.

Finally, several studies have tried to leverage [\ion{C}{ii}] 158-$\mu$m and [\ion{N}{ii}] 205-$\mu$m lines, in combination with radiative transfer modelling. Using this setup, \citet{Nagao2012} inferred a solar-like metallicity ($\log{Z/Z_\odot}=0.0\pm0.3$) in the $z=4.76$ galaxy ALESS 73.1 (see also \citealt{DeBreuck2014} who infer $Z=0.6-3.0$~$Z_\odot$). Similarly, \citet{deBreuck2019} inferred $Z=0.3-1.3$~$Z_\odot$ in a $z=4.2$ lensed DSFG SPT0418-47. Ultimately, our understanding of the metallicity distribution in the DSFG population will soon be revolutionised by the near- and mid-infrared spectroscopy with the \textit{James Webb Space Telescope}.

\begin{figure}
\center
\includegraphics[width = 0.45\textwidth]{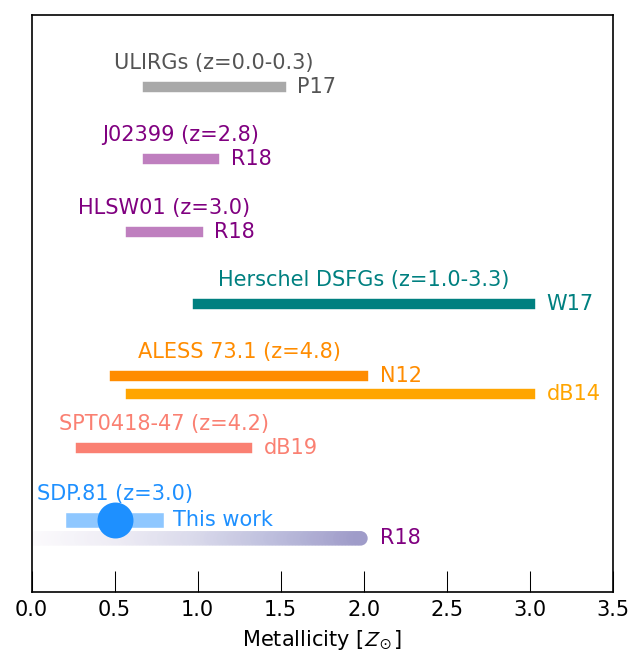}
\caption{Metallicity estimates and ranges for $z\geq1$ DSFGs from \citet[N12]{Nagao2012}, \citet[dB14]{DeBreuck2014}, \citet[W17]{Wardlow2017}, \citet[R18]{Rigopoulou2018}, \citet[dB19]{deBreuck2019}, and ULIRGs from \citet[P17]{Pereira2017}. For SDP.81, we show our best-fit solution $Z=0.5$~$Z_\odot$ (circle); we also include a tentative uncertainty of $\pm$0.25~$Z_\odot$. The sub-solar metallicity in SDP.81 is consistent with the upper limit of $\leq2~Z_\odot$ from \citet{Rigopoulou2018}, and comparable to the lower range of metallicity estimates in other high-z DSFGs. \label{fig:metallicity}}
\end{figure}

As illustrated in Fig.~\ref{fig:metallicity}, the $Z=0.5~Z_\odot$ metallicity inferred in SDP.81 falls below the metallicities derived by \citet{Wardlow2017}, but within the range inferred for ALESS~73.1 and SPT0418-47. Further support for the sub-solar metallicity in SDP.81 comes from HCO$^+$/HCN surveys of nearby star-forming regions: for example, the Large Magellanic Cloud (LMC, $Z\approx0.5$~$Z_\odot$) has HCO$^+$(4--3)/HCN(4--3) ratios $\geq$1 throughout most of its volume, with some sub-regions of 30~Doradus and N159W star-forming regions having HCO$+$/HCN$\geq$5 \citep{Anderson2014, Galametz2020}.

We further consider what the impacts of less-than-solar metallicities might be when inferring physical properties of DSFGs. In particular, one potential issue is related to inferring the molecular gas masses from low-$J$ CO emission. This conversion is dependent on the gas-phase metallicity: the CO-to-H$_2$ conversion factor $\alpha_\mathrm{CO}$ can increase rapidly with falling metallicity (see \citealt{Bolatto2013}); depending on the model, $\alpha_\mathrm{CO}$ increases by a factor of 2 to 5 \citep{Israel1997, Narayanan2012}. The value of $\alpha_\mathrm{CO}$ in high-redshift DSFGs remains controversial, with inferred values ranging from $\sim0.8$ (e.g. \citealt{Calistro2018, Frias2023}) to $\sim$6;  in fact, it is likely that DSFGs span a wide range of $\alpha_\mathrm{CO}$ \citep{Harrington2021}. However, as the bulk of DSFG studies assumed dusty galaxies to have (super)solar metallicity and $\alpha_\mathrm{CO}$ between 0.8 and $\approx$4, assuming a single $\alpha_\mathrm{CO}$ value for the DSFG population might cause a significant underestimation of gas masses in $Z\leq1\,Z_\odot$ DSFGs.

Finally, we note that in our PDR models, changing metallicity values merely ends up scaling the C, N, O abundances with respect to the solar values. However, this is not strictly physical, as the relative elemental abundances can vary with metallicity and the stellar population of the galaxy. For example, the empirical model of \citet{Dopita2016} (based on Galactic data) gives an N/O abundance ratio of $\approx$0.06 at 0.5~$Z_\odot$ and 0.1 for 1.0~$Z_\odot$. Similarly, the different production and destruction mechanisms for C, N, and O result in a strong variation of their relative abundances as the initial stellar population ages (e.g. \citealt{Maiolino2019}) with O/N and O/C abundance ratios decreasing as the starburst ages. The overabundance of O in young starbursts will be even more pronounced for a top-heavy stellar IMF, as has been claimed for DSFGs (e.g. \citealt{Zhang2018, Zhang2018b}). However, such a detailed analysis is beyond the scope of this paper.

\section{Conclusions}

We present deep ALMA observations of the mid-$J$ HCN, HCO$^+$, and HNC emission in SDP.81, a well-studied $z\sim3$ lensed dusty galaxy. Combining multi-epoch imaging, we obtained a robust detection of the HCO$^+$(4--3) emission, the third reported detection of this line in a high-redshift dusty galaxy.
The upper limits on the HCN(4--3)/CO(1--0) and HCN(4--3)/FIR ratios in SDP.81 are consistent with upper limits derived from HCN(1--0) observations from \citet{Rybak2022a}.

The simultaneous non-detections of the HCN(4--3) and HNC(4--3) lines imply a significantly elevated HCO$^+$/HCN luminosity ratio, making SDP.81 an outlier among extragalactic sources. Using a grid of PDR models, we find that the HCO$^+$, HCN, HNC, and CO observations of SDP.81 are consistent with a low amount of mechanical heating (0-10\% of the total energy input). This contradicts recent estimates based on high-$J$ CO emission studies of high-z DSFGs \citep{Harrington2021, Riechers2021}; however, the HCN/HCO$^+$/HNC lines are expected to be more direct tracers of mechanical heating \citep{Loenen2008, Kazandjian2015}. 

Our PDR modelling also indicates that SDP.81 has a sub-solar metallicity ($Z=0.5$~$Z_\odot$). This result is lower than typically assumed for high-redshift dusty galaxies, but within the range spanned by analyses of fine-structure lines for other redshift 2--5 sources.

We stress that the sub-solar metallicity in SDP.81 might not be representative of the DSFG population in general, as observations of mid-$J$ HCN and HCO$^+$ lines in high-z galaxies generally imply HCO$^+$/HCN ratios closer to one (\citealt{Bethermin2018, Canameras2021}, Rybak et al., in prep.). Nevertheless, if a fraction of DSFG have sub-solar gas-phase metallicity, this might have implications for the systematics of inferring their gas masses via the $\alpha_\mathrm{CO}$ conversion factor, among other processes.

\begin{acknowledgements}

The authors thank Z.~Zhang for sharing the data from \citet{Zhang2014} and providing valuable insights into the local HCN/HCO$^+$(4--3) observations, and K. Butler for sharing her CO(9--8) observations.\\

This paper makes use of the following ALMA data: \#2016.1.00663.S, \#2017.1.01694.S, and \#2018.1.00747.S.
ALMA is a partnership of ESO (representing its member states),
NSF (USA) and NINS (Japan), together with NRC (Canada), MOST and ASIAA (Taiwan), and KASI (Republic of Korea), in cooperation with the Republic of Chile. The Joint ALMA Observatory is operated by ESO, AUI/NRAO and NAOJ.\\
Based on observations collected at the European Organisation for Astronomical Research in the Southern Hemisphere under ESO programme(s) 294.B-5042(A).

The authors acknowledge assistance from Allegro, the European ALMA Regional Center node in the Netherlands.\\

M. R. is supported by the NWO Veni project "\textit{Under the lens}" (VI.Veni.202.225). J. H. acknowledges support of the VIDI research programme with project number 639.042.611, which is (partly) financed by the Netherlands Organisation for Scientific Research (NWO). S. V. acknowledges support from the European Research Council (ERC) grant MOPPEX ERC-833460. This work is based on the research supported in part by the National Research Foundation of South Africa (Grant Numbers: 128943).

\end{acknowledgements}

\bibliographystyle{aa}
\bibliography{references}

\begin{appendix}

\section{Environment of the lensing galaxy}
\label{app_A}

In addition to the $z=3.042$ lensed DSFG, our ALMA Band~3 observations provide additional insights into the environment of the lensing galaxy. In particular, we find an indication of a foreground CO(0--1) absorption and a nearby, CO(1--0) luminous galaxy.

\subsection{CO(0--1) absorption from the foreground galaxy}

In addition, the emission from the SDP.81 itself, we find a pronounced negative feature around 87.9~GHz, with a total flux of 0.41$\pm$0.05~Jy km~s$^{-1}$ (Fig.~\ref{fig:alma_spectrum}). This feature is not associated with the lensing arcs and is likely originating in the foreground galaxy. A potential candidate would be a CO(0--1) absorption at $z=0.310$.In such case, the absorbing gas would be offset by $\approx$30,000 km s$^{-1}$ from the systemic redshift $z_L$=0.299\footnote{We confirm the redshift of the lensing galaxy as 0.299$\pm$0.001 based on the absorption lines seen in the MUSE spectroscopy (see Appendix~A)}. 

We hypothesise that this might be absorption due to a foreground structure. As we show in Appendix~A, we also detected CO(1--0) emission in a $z=0.301$ source to the south-east of the lensing galaxy. Indeed, CO(0--1) absorption has been detected in several nearby and intermediate-redshift galaxies, ranging from Centaurus A \citep{Israel1991} to the $z=0.88$ foreground galaxy of PKS 1830-211 \citep{Merin1997}.

\begin{figure}[h]
\centering
    \includegraphics[width=0.33\textwidth]{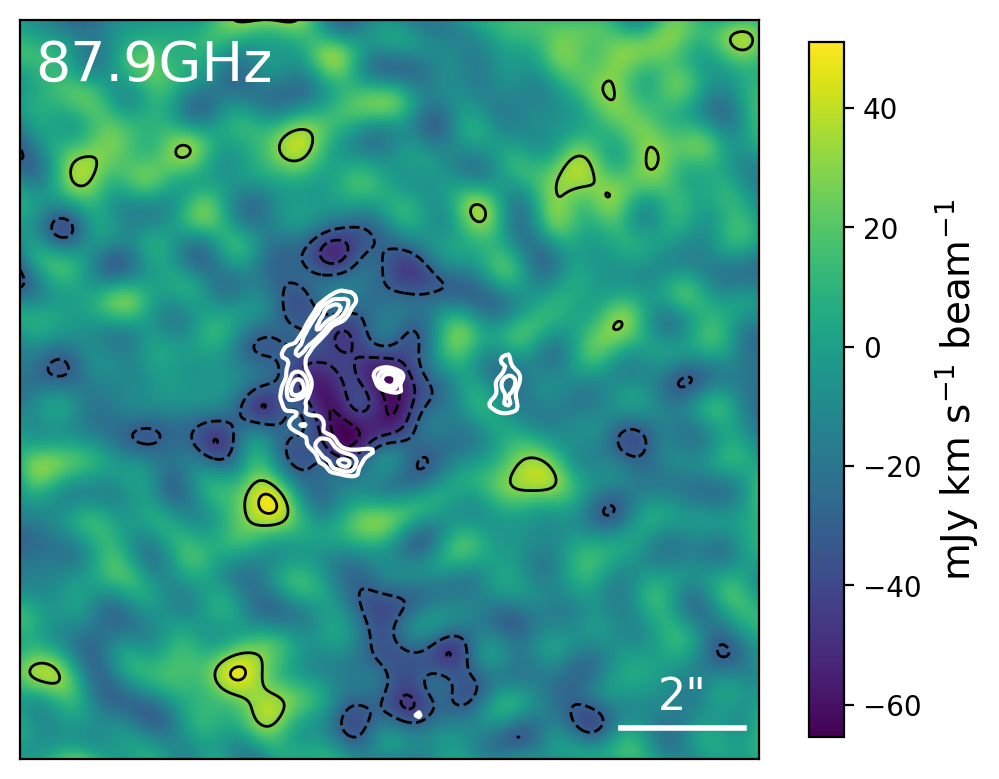}
     
    \caption{Image of the absorption feature at 87.9~GHz, made from the continuum-subtracted data using a 250-k$\lambda$ uv-distance cut. The black contours start at $\pm$2$\sigma$ and increase in steps of $1\sigma$, white contours indicate the Band~3 continuum. The absorption is not associated with the lensing arcs and is most likely due to a foreground CO(0--1) absorber at $z\approx0.31$.}
    \label{fig:co10_absorption}
\end{figure}

\subsection{A serendipitous discovery of CO(1--0) emission from a companion of the SDP.81 lensing galaxy}
\begin{figure}[h]
\centering
    \includegraphics[width=0.49\textwidth]{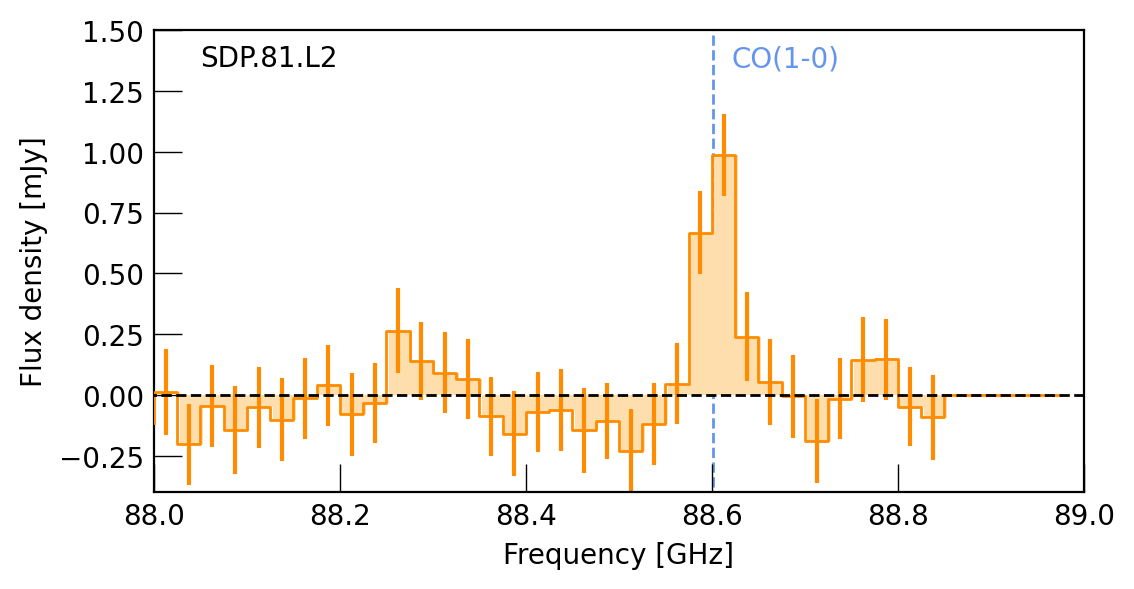}\\
     
    \caption{CO(1--0) detection in a $z\sim0.301$ foreground galaxy, extracted from a circular aperture (4-arcsec diameter).}
    \label{fig:companion}
\end{figure}

In addition to SDP.81, we detect line emission from another source in the field, offset by $\approx$12~arcsec ($\approx$55~kpc) to the south-east (J2000~09:03:12.05 00:38:51.5). The line is centred at 88.62~GHz, a total line flux of 0.9~Jy km s$^{-1}$ and with a FWZI width of 210 km s$^{-1}$ (Fig.~\ref{fig:companion}). This corresponds to the expected frequency of the CO(1--0) at the redshift $z=0.301$, offset by $\sim$2000 km s$^{-1}$ from the lensing galaxy ($z_L =0.299$).

\begin{figure}[h]
\centering
    \includegraphics[width=0.49\textwidth]{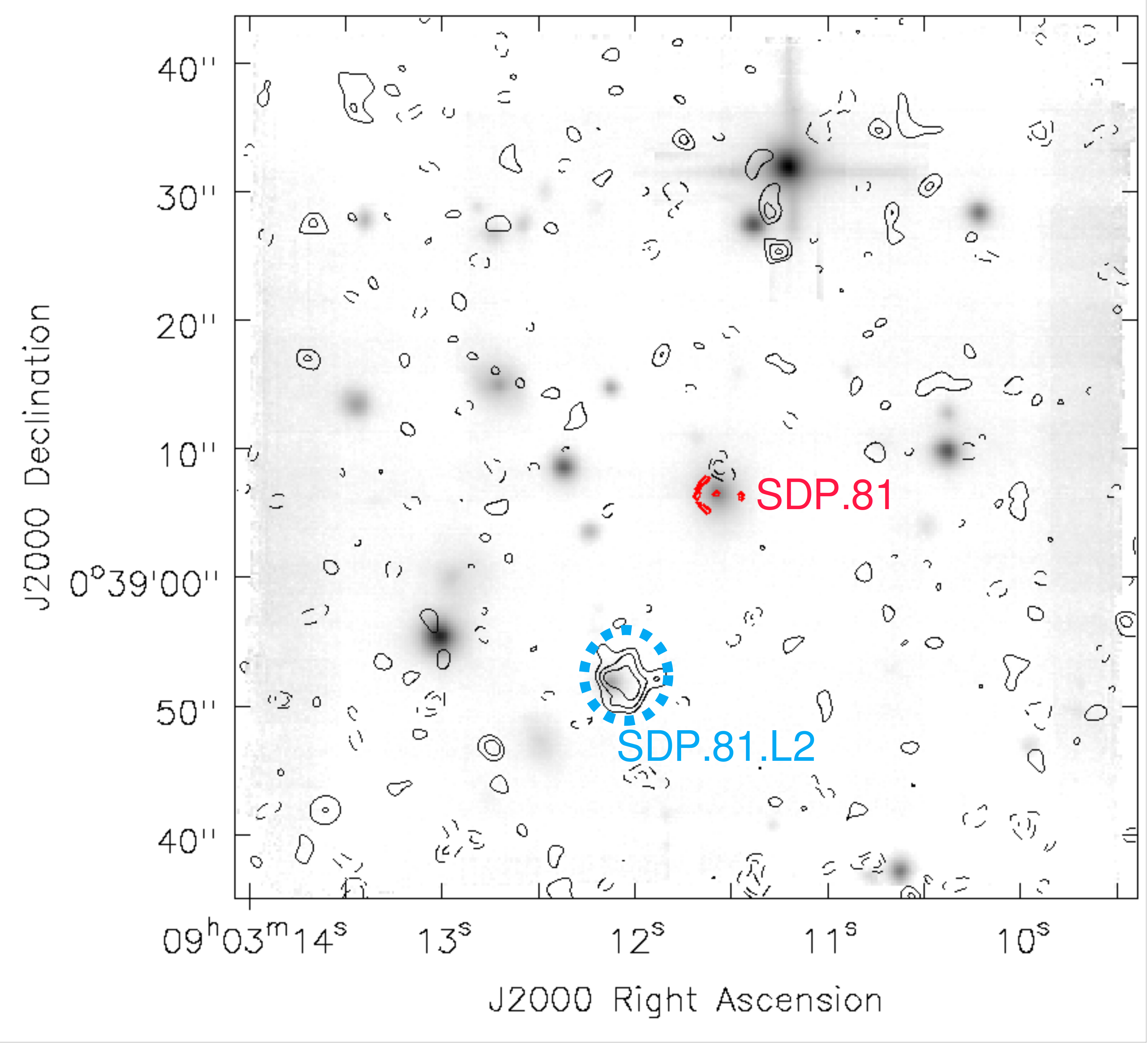}\\
     
    \caption{Overlay of the MUSE 465 - 930~nm continuum (greyscale), ALMA CO(1--0) emission (black contours, integrated over 88.55 - 88.65~GHz), and ALMA Band~3 continuum (red contours). The CO(1--0) contours start at $2\sigma$ and increase in steps of 1$\sigma$. The peak CO(1--0) S/N is 4.9$\sigma$. The CO(1--0) emission in the blue circle is co-spatial with an optically bright MUSE continuum source.}
    \label{fig:companion_MUSE_ALMA}
\end{figure}

Figure~\ref{fig:companion_MUSE_ALMA} shows a comparison of the ALMA CO(1--0) emission and the optical VLT-MUSE imaging of the SDP.81 field (Programme ID: 294.B-5042(A), PI: Gavazzi). The MUSE continuum image (greyscale) shows an optical counterpart to the CO(1--0) emission. Our examination of the MUSE spectrum (not shown here) shows prominent emission lines (H$\alpha$, [\ion{N}{ii}], [\ion{O}{iii}]) redshifted to $z\sim0.3$. This confirms that the observed emission is indeed CO(1--0) from a foreground galaxy.

We estimate the CO(1--0) luminosity as $L_\mathrm{CO(1-0)}=2\times10^6$~$L_
\odot$, $L'_\mathrm{CO(1-0)}=4.5\times10^9$~K km s$^{-1}$ pc$^{2}$. 
Fitting a Gaussian profile to the observed emission using the \texttt{imfit} task yields a source size of $\approx$ 3.9" $\times$ 2.8".
Therefore, we hypothesise that this is a gas-rich counterpart to the lensing elliptical galaxy, with $M_
\mathrm{gas}\simeq2\times10^{10}$~$M_\odot$ (assuming a Milky-Way CO-to-H$_2$ conversion factor $\alpha_\mathrm{CO}=4.3$ $M_\odot$/(K km s$^{-1}$ pc$^2$), \citealt{Bolatto2013}, and including the contribution of helium). The gas mass of this potential companion is comparable to the stellar mass of the main lensing galaxy ($M_\star\simeq2\times10^{10}$~$M_\odot$, \citealt{Negrello2014}). Such a massive companion might have an influence on the details of the lens model, for example, for the analysis of substructure in the lensing galaxy \citep{Hezaveh2016}.

\end{appendix}
   
\end{document}